\documentclass[aps,prd,twocolumn,nofootinbib,showpacs,superscriptaddress,groupedaddress]{revtex4}  
\usepackage{graphicx}  
\usepackage{dcolumn}   

\usepackage{amsmath}
\usepackage{amsfonts}
\usepackage{epsfig}
\usepackage{slashed}
\usepackage{braket}

\usepackage{comment}
\usepackage{cases}
\def\KD{K\"{a}hler-Dirac }

\def\bec{\begin{center}}
\def\eec{\end{center}}
\def\beq{\begin{equation}}
\def\eeq{\end{equation}}
\def\bea{\begin{eqnarray}}
\def\eea{\end{eqnarray}}

\begin{document}

\title{Chiral Lattice Fermions From Staggered Fields}
\author{Simon Catterall}
\affiliation{Department of Physics, Syracuse University, Syracuse, NY 13244, USA }
\date{\today}

\begin{abstract}
We describe a proposal for constructing a lattice 
theory that we argue may be capable of yielding free Weyl fermions in the continuum limit. The model employs
reduced staggered fermions and uses site parity dependent 
Yukawa interactions of Fidkowski-Kitaev type to gap a subset of the lattice fermions without breaking
symmetries. The possibility for such symmetric mass generation is tied to the cancellation of certain discrete
anomalies arising in the continuum limit. The latter place strong constraints on the number of lattice
fermions -- constraints that are satisfied by this model.
We present numerical results for the model in two dimensions which support this scenario. \end{abstract}


\maketitle

\section{Introduction}

It has long been a goal of lattice field theory to be able to describe continuum chiral
gauge theories. All of the standard local lattice fermion prescriptions; Wilson, 
staggered, domain wall and overlap appear only capable of describing vector-like theories. The central reason for this is well known - the Nielsen-Ninomiya theorem asserts that a wide class of fermion discretizations with exact chiral symmetry will necessarily contain equal numbers of left and right handed fields \cite{Nielsen:1981hk}.

Two paths have typically been followed to try and evade this theorem. In domain wall approaches one starts from a lattice in five dimensions
and then introduces a domain wall in the extra dimension which binds one fermion of fixed four dimensional chirality. On a finite lattice one must necessarily introduce
an anti-domain wall binding a fermion of opposite chirality. Unfortunately in the presence of a fluctuating gauge field these modes become coupled once more and  the continuum limit describes Dirac fermions again. Recently Grabowska and Kaplan proposed a modification of the domain wall prescription that used a deformation of the fermion kernel to decrease
the coupling to the mirror fermions located on the anti-domain wall  \cite{Grabowska:2015qpk}. However, it is unclear whether this proposal will be sufficient at the
non-perturbative level.

A second way to proceed is to start with a 
vector-like theory directly in four dimensions and attempt to give cut-off scale masses to
states of one chirality using strong multifermion interactions. 
Perhaps the first example of such a mirror model was given by 
Eichten and Preskill  \cite{Eichten:1985ft}. 
The early numerical work to test this idea made use of Wilson and staggered lattice fermions \cite{Bock:1993iv,Bock:1990cx,Hasenfratz:1991it,Hasenfratz:1988vc,Lee:1989xq} and 
appeared to invalidate the approach -  to generate 
large mirror masses required large four fermion or Yukawa couplings
and typically this resulted in the formation of symmetry breaking condensates
coupling left and right handed states via Dirac mass terms \cite{Golterman:1992yha}.  

More recently this approach was revived 
for lattice fermion actions with superior chiral properties - in a series of
papers Poppitz et al. have investigated models using overlap fermions \cite{Poppitz:2009gt,Poppitz:2010at,Chen:2012di} while
a gauge invariant path integral measure for overlap chiral fermions in $SO(10)$ was constructed in \cite{Kikukawa:2017ngf}.
An earlier proposal combining domain wall fermions
and appropriate four fermion interactions was made by Creutz et al in \cite{Creutz:1996xc}.
However, again, the overall conclusion
of these studies was that it was difficult, if not impossible, to decouple 
the chiralities in the continuum limit.\footnote{A possible exception to this was L\"{u}scher's formal 
construction of a path integral for $U(1)$ chiral gauge theory in \cite{Luscher:1998du} }

However, in recent years, a series of developments in condensed matter physics have provided new insights into the problem. One of the key new ingredients has been the discovery of models capable of symmetric mass generation. This 
field was launched by the
seminal paper of Kitaev and Fidkowski (FK) \cite{Fidkowski:2009dba} who showed that it was possible to design a four fermion interaction that was capable of generating masses for
precisely eight one dimensional Majorana modes without generating symmetry
breaking fermion condensates. Subsequent work
generalized this to higher dimensions 
finding that sixteen Majorana fermions are needed in three
and four (spacetime) dimensions \cite{Ryu:2012he,Qi:2013dsa,You:2014vea,Tachikawa:2018njr}. 

It is now understood that the appearance of specific numbers of fermions in
theories where symmetric mass generation is possible, is
tied to the cancellation of certain discrete anomalies  \cite{Kapustin:2014dxa,Garcia-Etxebarria:2018ajm}. 
Indeed, one way to understand the observation of 
symmetry broken phases in some of the earlier work
on lattice four fermion theories is that they arise
as a consequence of the failure to
cancel off these discrete anomalies. To replicate a non-zero anomaly in the UV requires massless states
in the IR which can arise as Goldstone modes arising after the spontaneous breaking
of symmetries via fermion condensates.

The phenomenon of symmetric mass generation has been seen in 
recent numerical studies of vector-like lattice models 
in dimensions from two to four 
\cite{Ayyar:2014eua,Ayyar:2015lrd,Catterall:2015zua,Catterall:2016dzf,Butt:2018nkn,Ayyar:2016lxq,Ayyar:2017qii,You:2017ltx}. While massive symmetric phases
had been observed in early lattice studies of Higgs-Yukawa theories, they were typically separated from the weak coupling regime by regions where lattice symmetries were spontaneously broken by the formation of bilinear condensates and the massive symmetric phases were 
interpreted as lattice artifacts \cite{Bock:1993iv,Bock:1990cx,Hasenfratz:1991it,Hasenfratz:1988vc,Lee:1989xq}.  In the more recent work these massive symmetric
phases in theories can be directly connected to the massless phase via a continuous phase transition thereby allowing 
for a continuum limit. One crucial difference between the new
work and these older studies is the use of reduced staggered
fermions which carry half the number of degrees of freedom of regular staggered
fermions and can be thought of as lattice analogs of continuum Majorana fermions. They 
will form a key ingredient in this proposal for constructing chiral lattice theories.

It was realized some years ago by Xu, Wen, and others in the condensed matter community
that symmetric mass generation might allow one to construct anomaly free chiral lattice gauge theories and several
proposals have been made \cite{You:2014vea,You:2014oaa,Wang:2018cai,Wang:2018ugf}. However, 
these models use Hamiltonian formulations and continuum topological arguments to make 
the case for symmetric mass generation
The construction in this paper aims to furnish an explicit Euclidean lattice path integral
in which single component relativistic lattice fermions can be gapped by a FK type interaction
thereby producing a low energy theory which can be shown to produce chiral fermions in the continuum limit.

We start our discussion with a quick review of reduced staggered fermions and how they may be given a mass without breaking symmetries 
using a four fermion interaction. However, this construction yields a continuum limit describing massive Dirac fermions.
To target a chiral theory 
requires different field content and interactions. We argue that the requirement that the lattice model
yield the correct number of massless Majorana fermions in the continuum limit
suggests a specific field content and an interaction of Fidkowski-Kitaev type. 

We then discuss numerical results from simulations of the simplest model in two
dimensions and provide evidence that indeed the interactions we propose are
capable of decoupling the 
relevant lattice modes without the formation of bilinear
condensates.
We summarize our conclusions and discuss open questions in the final section of the paper.

\section{Reduced Staggered Fermions and Symmetric Mass Generation}

The usual staggered fermion action in $D$ dimensions is easily arrived at by spin diagonalizing the naive fermion
action on a hypercubic lattice and takes the form \cite{Golterman:1984cy}  
\begin{equation}
    S=\sum_{x,\mu}\eta_\mu(x) \overline{\chi}(x)D^S_\mu\chi(x)+\sum_x m\overline{\chi}(x)\chi(x)
\end{equation}
where $\eta_\mu(x)=\left(-1\right)^{\sum_{i=0}^{\mu-1}x_i}$ are the usual staggered fermion
phases and the symmetric difference is given by
\begin{equation}
    D^S_\mu \chi^a(x)=\frac{1}{2}\left(\chi^a(x+\mu)-\chi^a(x-\mu)\right)
\end{equation}
If $m=0$ a further reduction is possible by keeping only one (single component) fermion at each lattice site. Explicitly we introduce the projectors $P_\pm$ defined by
\begin{equation}
    P_\pm=\frac{1}{2}\left(1\pm \epsilon(x)\right)
\end{equation}
where the site parity is given by $\epsilon(x)=\left(-1\right)^{\sum_{\mu=0}^{D-1} x_\mu}$.
The lattice action decomposes into
\begin{equation}
    S=\sum_{x,\mu} \eta_\mu(x)\left(\overline{\chi}_+(x)D^S_\mu\chi_-(x)+
    \overline{\chi}_-(x)D^S_\mu\chi_+(x)\right)
\end{equation}
where $P_+\chi=\chi_+$ etc.
The reduction we need corresponds to, for example, retaining only the fields
$P_+\chi$ and $P_-\overline{\chi}$. This results in the reduced staggered fermion action whose continuum limit corresponds to $2^{D/2-1}$ Dirac fermions or equivalently
$2^{D/2}$ Majorana fermions \cite{vandenDoel:1983mf}. 
\begin{equation}
S=\sum_{x,\mu} \chi^a(x)\eta_\mu(x)D^S_\mu\chi^a(x)
\end{equation}
where we have relabeled $\overline{\chi}\to \chi$ on odd parity lattice sites. 
This action is invariant under both discrete rotations and a staggered shift symmetry:
\begin{equation}
    \chi(x)\to \xi_\mu(x)\chi(x+\mu)
\end{equation}
where $\xi_\mu=\left(-1\right)^{\sum_{i=\mu+1}^D x_i}$ \cite{vandenDoel:1983mf}.
The reduced staggered action is also invariant under a $U(1)$ symmetry 
\begin{equation}
    \chi(x)\to e^{i\alpha\epsilon(x)}\chi(x)\\
\end{equation}
While the single flavor theory does not allow for a mass term this can be remedied
by introducing multiple flavors of reduced staggered field. To avoid fermion bilinear terms the simplest model 
requires four reduced staggered fields transforming under
an $SU(4)$ global symmetry
\begin{align}
   S&=\sum_{x,\mu} \chi^a(x)\eta_\mu(x)D^S_\mu\chi^a(x)\\
   &-\frac{G^2}{8}\sum_x\epsilon_{abcd} \chi^a(x)\chi^b(x)\chi^c(x)\chi^d(x)\nonumber
\end{align}
Notice that the four fermion terms break the $U(1)$ symmetry down to $Z_4$ which acts on the lattice fermions
as
\beq
\chi(x)\to i\epsilon(x)\chi(x)\eeq
This  combination of shift, $Z_4$ and $SU(4)$ symmetries
forbid any fermion bilinear operator from appearing in the quantum effective action.
\begin{figure}[htb]\centering
    \includegraphics[width=0.4\textwidth]{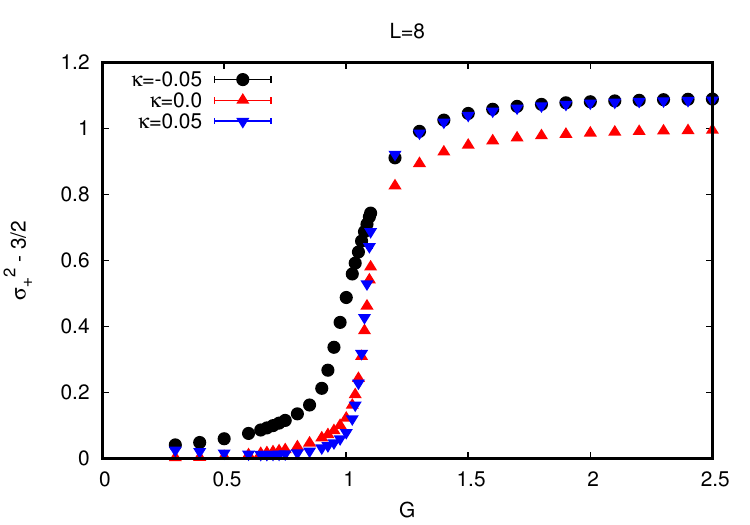}
    \caption{Four fermion condensate vs $G$}
    \label{fig1}
\end{figure}
One intuitive way to
understand how a 
fermion mass can arise in these models is to rewrite the four fermion operator as
\begin{equation}
    \epsilon_{abcd}\chi^a(x)\chi^b(x)\chi^c(x)\chi^d(x)=\Omega_a(x)\chi^a(x)
\end{equation}
corresponding to a bilinear mass term formed by pairing an elementary fermion with a composite fermion $\Omega_a(x)=\epsilon_{abcd}\chi^b(x)\chi^c(x)\chi^d(x)$.
It is easy to see that a condensate of this four fermion operator
arises in the strong coupling limit $G\to\infty$ which can hence be interpreted as generating
a fermion mass in that same limit.
Since one expects massless fermions
at $G=0$ there must be at least one phase transition separating the
massless and massive symmetric regimes.\footnote{The exception to this is two dimensions where the four
fermion coupling is asymptotically free and one sees \cite{Ayyar:2017qii} a
single symmetric gapped phase for all non-zero lattice couplings.}

Evidence of a continuous transition separating these phases has been seen in both three and
four dimensions \cite{Ayyar:2014eua,Ayyar:2015lrd,Ayyar:2016lxq, Catterall:2015zua,Catterall:2016dzf,Butt:2018nkn}.
To underline these conclusions we include a plot in fig.~\ref{fig1} of
the four fermion condensate in the four dimensional model
taken from our
earlier paper \cite{Butt:2018nkn}. The 
rapid increase of the condensate close to $G\sim 1$ is indicative of such a phase transition which is borne out by further analysis as described
in \cite{Butt:2018nkn}. In this plot the coupling $\kappa$ that appears
is the coefficient of an additional
scalar kinetic term which must be tuned in the four dimensional theory to see this direct transition
between massless and massive symmetric phases. The appearance of this
additional operator is natural within a RG framework since the scalar kinetic
first becomes marginally relevant in four dimensions.

These results encourage the
belief that it is possible to achieve symmetric mass generation in relativistic
lattice theories based on reduced staggered
fermions even in 
four dimensions. However,
since reduced staggered fields yield Dirac fermions in the
continuum limit the mechanism described above is only capable
of generating mass in vector-like theories. To target a chiral
theory  we need to modify the quartic interactions in a manner that
subjects only a subset of the reduced staggered field to quartic
interactions. In a staggered theory there is a natural
way to do this by dividing the lattice field into its even and odd site parity
components. One is thus led to consider models in which Yukawa
interactions are applied only to say even parity fields. We discuss the necessary structure for these
interactions in the next section.

\section{Modified Quartic Interactions }

We consider an action of the form:
\begin{align}
\label{kitaevint}
S&=\sum_{x,\mu} \eta_\mu(x)\chi^a(x)D^S_\mu\chi^a(x)\\\nonumber
&-\sum_x
\left(GP_+ +gP_-\right)
\left[\chi^T(x)\Gamma_A\chi(x)\right]^2
\end{align}
where $\Gamma_A$ are a set of antisymmetric matrices. As we will show in the next section the requirement 
that we target chiral fermions in the continuum limit requires the staggered fields to transform in a real
representation of any underlying symmetry group. This will also guarantee the absence of additional
perturbative and 't Hooft anomalies. 
One simple way to ensure this constraint 
it to look for real representations of a Euclidean rotation group. The smallest such group  corresponds
to ${\rm Spin}(7)$ which possesses a real, eight dimensional  spinor representation. In this
case the $\Gamma_A$ appearing in eqn.~\ref{kitaevint} are taken to be the
pure imaginary, antisymmetric Dirac matrices appropriate for this representation.  They are given by
\begin{equation}
    \Gamma_A=\left(\sigma^{123},\sigma^{203},\sigma^{323},\sigma^{211},\sigma^{021},\sigma^{231},\sigma^{002}\right)
\end{equation}
where the notation indicates that the Dirac matrices $\Gamma_A$ are built from tensor
products of Pauli matrices. The resultant term is precisely the Fikdowski-Kitaev 
interaction which is known to be capable of generating mass for
Majorana fermions in one dimension \cite{Fidkowski:2009dba}.

In practice we generate this term by
coupling the bilinear $\chi^T\Gamma_A\chi$ to an auxiliary scalar $\sigma_A(x)$ with a simple
gaussian action $\sum_x \frac{1}{2}\sigma_A^2$.
Notice that if these scalar fields are taken constant the Yukawa interaction just becomes
a Majorana mass term for $8$ real staggered fermions. In principle these eight real fermions can be organized into four
complex staggered fields so that the theory will be invariant under the same $Z_4$ symmetry seen in
the previous model targeting vector-like theories.  In the appendix we
make another argument for the appearance 
of such a $Z_4$ symmetry based on a novel anomaly of free staggered fermions propagating on lattices with
non-trivial topologies.

A crucial feature of  the interaction we employ is that the associated Yukawa
coupling depends on the parity of a lattice site.
We will take $G$ to be large to drive
symmetric mass generation in the even site parity sector while $g$ is kept small
and serves merely to regulate a zero mode that appears in the odd parity sector at $g=0$
corresponding to the shift symmetry
\beq
\chi_-(x)\to \chi_-(x)+\alpha_-\eeq

In the next section we will show that the remaining light fermions after gapping are capable of being
organized in the continuum limit into 2 pairs of chiral
fermions so that the final 
theory will contain sixteen Weyl fermions.
This result is in accordance with the vanishing of a discrete spin-$Z_4$ anomaly for systems of Weyl fermions 
corresponding to the transformations \cite{Tachikawa:2018njr,Garcia-Etxebarria:2018ajm}:
\begin{align}
\psi_L&\to -i\psi_L\\
\psi_R&\to +i\psi_R
\end{align}
and takes the form
\begin{equation}
    \nu_4=n_+-n_-\quad {\rm mod}\;16
\end{equation}
where $n_\pm$ denote the number of left and right handed Weyl fields.

The Yukawa interaction we have
described can be reduced to the subgroups ${\rm Spin}(N)$ for $2\le N<7$
by truncating the index $A$ to  run from $1\ldots N$ as described in \cite{You:2014vea}.
The resultant theories with reduced symmetry still satisfy the anomaly cancellation condition and hence should still 
still be capable of symmetric mass generation.
Indeed, in the condensed matter literature the key 
feature which makes
this mechanism possible is the 
non-degeneracy of the ground state of
the FK interaction which continues
to hold even in the reduced symmetry cases. 
Systems with a unique ground state cannot undergo
spontaneous symmetry breaking and hence offer the possibility of symmetric mass generation.

\section{\label{chiral}Chirality and the Continuum Limit}

In this section we discuss the continuum limit and show how the gapped theory can exhibit a chiral
spectrum. As a warm up let us first consider the free two dimensional theory.
We start by assembling the reduced staggered fields in a unit square of the lattice into a $2\times 2$ matrix
field $\Psi$ labeled by both spinor and flavor indices. If we employ the basis $\{I,\sigma_1,\sigma_2,\sigma_1\sigma_2\}$ it is given by
$$\Psi=$$
\vspace{-0.75cm}
\begin{equation}
\left[\begin{array}{cc}
(\chi_+(x)+i\chi_+(x+1+2))&(\chi_-(x+1)+i\chi_-(x+2))\\
(\chi_-(x+1)-i\chi_-(x+2))&(\chi_+(x)-i\chi_+(x+1+2))\end{array}\right]
\end{equation}
where the notation $x+1$ indicates the lattice site one step in the 1-direction from site $x$ etc and
we have
suppressed the ${\rm Spin}(7)$ indices for simplicity.
$\Psi$ is defined on a lattice with twice the lattice spacing and we added explicit $\pm$ subscripts to $\chi$
indicating the site parity for clarity. 
Notice that this is a chiral basis for the two dimensional Dirac matrices
and hence the upper row of this matrix contains right handed fields while the left handed fields are located in the
lower row.  

In addition, since we are employing real staggered fields  the matrix field $\Psi$
satisfies a reality condition $\psi^*=\sigma_1\psi\sigma_1$ corresponding to the fact that it depends on
only four real fields. This structure allows us to build two Majorana spinors from
$\Psi$ 
\begin{align}
\Psi_1=\left(\begin{array}{c}
\chi_-(x+1)+i\chi_-(x+2)\\
\chi_-(x+1)-i\chi_-(x+2)\end{array}\right)\\
\Psi_2=\left(\begin{array}{c}
\chi_+(x)+i\chi_+(x+1+2)\\
\chi_+(x)-i\chi_+(x+1+2)\end{array}\right)\end{align}
Notice that each of these Majorana spinors depend only on lattice fields of  fixed parity.
If we are successful in generating a cut-off scale mass for the $\chi_+$ modes we are
left with a single light continuum Majorana fermion  corresponding to $\Psi_1$. 
In the $g\to 0$ limit this becomes massless and is equivalent to a single left handed Weyl fermion.

There is a similar story in four dimensions where
we can build  a $4\times 4$ matrix field $\Psi$ from the staggered
fields in a unit hypercube \cite{Bock:1992yr, Golterman:1984cy}.\footnote{In the lattice
literature this is termed the spin-taste basis. It is equivalent to the K\"{a}hler-Dirac representation used in lattice susy constructions \cite{Catterall:2009it}.}  
\begin{equation}
    \Psi=\sum_{\{n_\mu=0,1\}}\chi(x+n)\gamma^{x+n}
\end{equation}
where $\gamma^{n_\mu}=\gamma_0^{n_0}\gamma_1^{n_1}\gamma_2^{n_2}\gamma_3^{n_3}$.
In a chiral basis it is easy to see that $\Psi$ has the block structure
\begin{equation}
    \Psi=\left(\begin{array}{cc}E&O^\prime\\
    O&E^\prime\end{array}\right)
\end{equation}
where the $2\times 2$ block matrices $E,E^\prime$ and $O,O^\prime$ contain 
only even and odd lattice site staggered
fields. As in two dimensions the use of a real staggered field implies that
$\Psi$ obeys a reality condition:
\begin{equation}
    \Psi=\gamma_2\Psi^*\gamma_2
\end{equation}
This in turn ensures that $O^\prime=-\sigma_2O^*\sigma_2$ and $E^\prime=\sigma_2 E^*\sigma_2$.
This condition ensures that the action can be written just in terms of the blocks $O$ and $E$ which depend only on
the 16 real single component lattice fermions in an elementary hypercube.

Let us assume that we are successful in removing the even parity states from the low energy spectrum
via the strong Yukawa interaction. If Lorentz invariance is restored in the continuum limit this would
leave behind a pair of Majorana spinors determined only by the odd parity lattice fields:
\begin{align}
\Psi_1=\left(\begin{array}{c}
-\sigma_2 O^*\sigma_2\\
O\end{array}\right)
\label{maj}\end{align}
Notice that a conventional Majorana spinor can be written
\begin{equation}
\left(\begin{array}{c}i\sigma_2\chi^*\\
\chi\end{array}\right)
\end{equation}
The form of eqn.~\ref{maj} thus suggests we adopt 
a generalized charge conjugation operation which acts both on the Lorentz
and flavor indices of a pair of Weyl spinors. Again, if these Majorana spinors remain massless in the continuum
limit they can be replaced by the pair of left Weyl fields contained in the block $O$.

Of course this analysis assumes we are able to give masses to the even parity fields while leaving the odd parity
fields massless and non-interacting. We now give an argument in support of this conjecture.
We start by rescaling the fields according to:
\begin{align}
\chi_+&\to \frac{1}{G}\chi_+&\chi_-\to \frac{1}{g}\chi_-\\\nonumber
\sigma_+^A&\to \frac{1}{G}\sigma_+^A&\sigma_-^A\to \frac{1}{g}\sigma_-^A
\label{rescaled}
\end{align}
The rescaled action (suppressing the ${\rm Spin}(7)$ indices) then reads:
\begin{align}
S&=\frac{1}{x}\sum_x
 \left[  
\chi_+ \eta_\mu D_\mu \, \chi_-  +
 y\chi_+\Gamma_A\chi_+\sigma_+^A +\right.\\
&\left.\frac{1}{y}\chi_-\Gamma_A\chi_-\sigma_-^A
+\frac{1}{2}y(\sigma_+^A)^2+\frac{1}{y}(\sigma_-^A)^2
\right]\nonumber
\end{align}
where $x=Gg$ and $y=\frac{g}{G}$.
The equations of motion for the staggered fields are then
\begin{align}
    \eta_\mu D_\mu \chi_-+y(\chi_+\Gamma_A\chi_+)\Gamma_A\chi_+&=0\\
    \eta_\mu D_\mu \chi_++\frac{1}{y}(\chi_-\Gamma_A\chi_-)\Gamma_A\chi_-&=0
\end{align}
In the limit $y\to 0$ we see that the odd site field $\chi_-$,
corresponding to the blocks $(O,O^\prime)$, is weakly coupled while the even
parity field $\chi_-$, represented by $(E,E^\prime)$, is strongly coupled.
In the next section we will provide numerical evidence that the
fields in $(E,E^\prime)$ are in fact gapped and decoupled from the long wavelength
modes given by the blocks $(O,O^\prime)$.  
\begin{figure}[htb]\centering
    \includegraphics[width=0.4\textwidth]{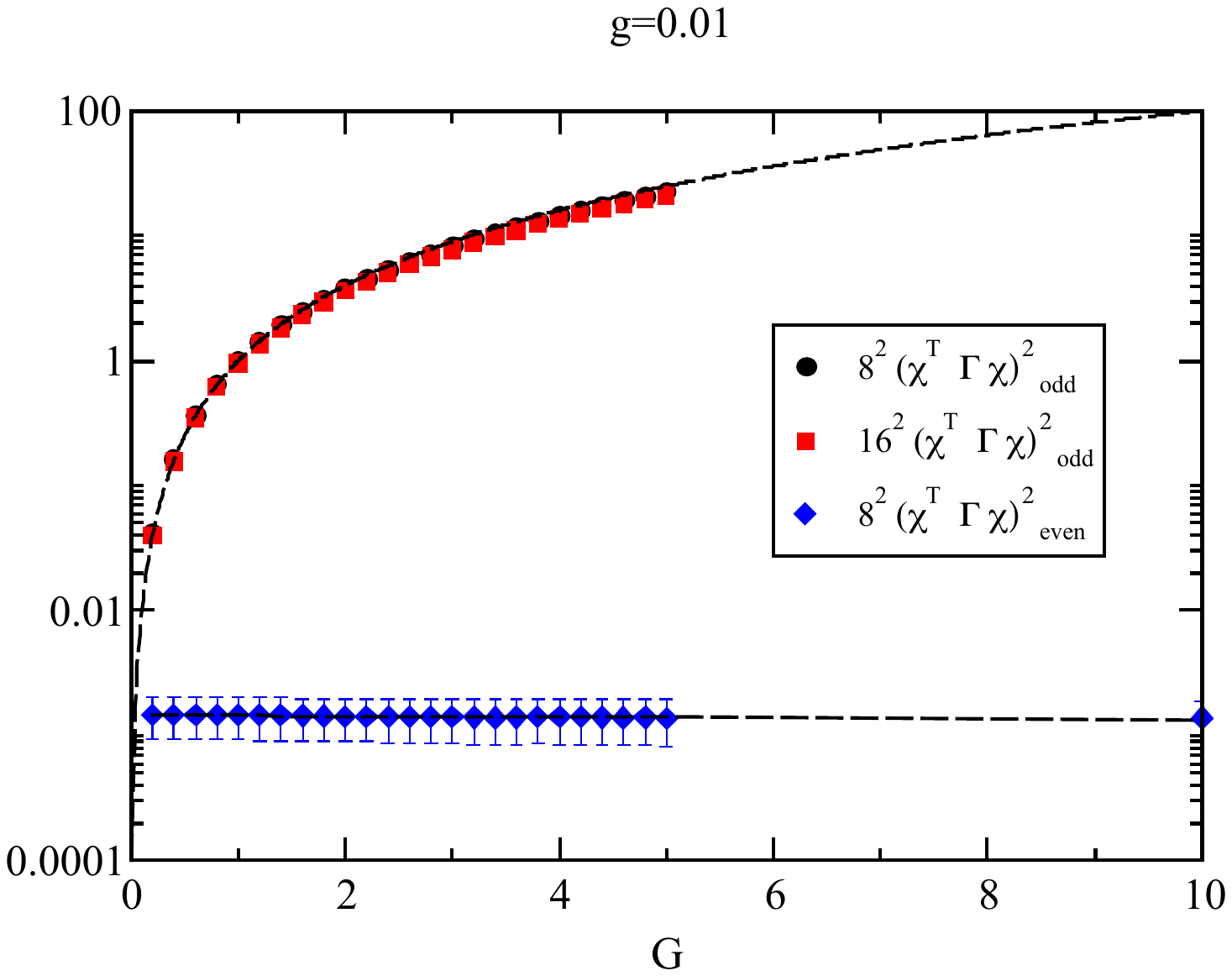}
    \caption{$<(\chi^T\Gamma\chi)^2>$ vs $G$ for $L=8,16$}
    \label{four}
\end{figure}
\section{\label{numerics}Evidence for Site Parity Dependent Mass Generation }

In this section we present preliminary numerical evidence that the model
is indeed capable of generating mass for the even site fermions without
generating symmetry breaking fermion bilinear condensates. To avoid the need for tuning
a scalar kinetic term results are presented only for two dimensions~\footnote{Work in 
four dimensions is in progress and will be reported in a separate publication}. However, both the
structure of the FK interaction, the argument for symmetric mass generation,
and appearance of chiral fermions in the continuum hold in both the two and four dimensional models.

We use a RHMC algorithm to simulate the system.
For more details of our numerical methods see appendix~\ref{rhmc}.
Measurements of the phase of the Pfaffian resulting from fermion integration
show an absence of a sign problem for all the ensembles presented in this paper.
\begin{figure}[htb]\centering
    \includegraphics[width=0.4\textwidth]{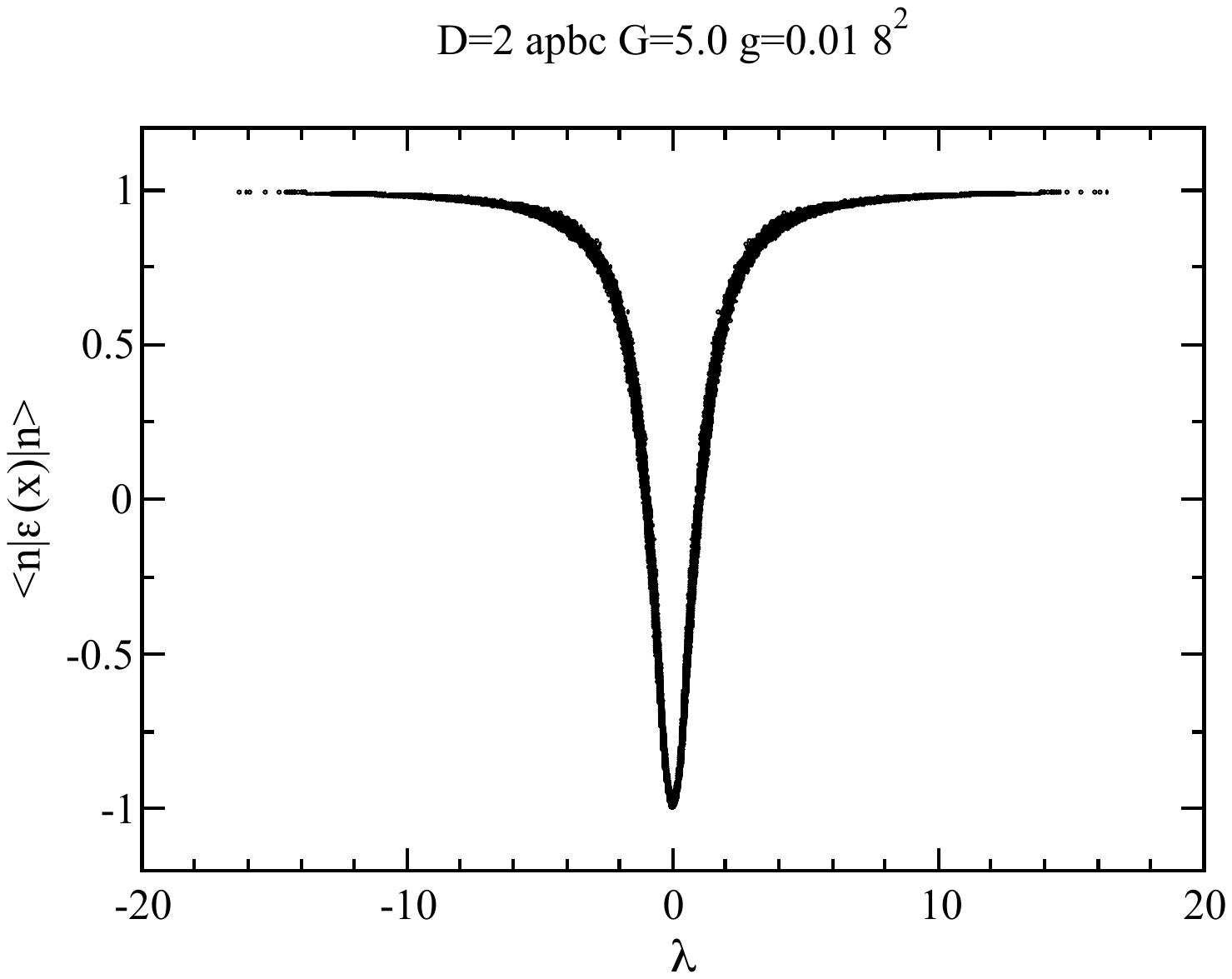}
    \caption{$<n\epsilon(x)n>$ vs $\lambda_n$ for $8^2$ lattice at $G=5.0$ and $g=0.01$}
    \label{eigen}
\end{figure}
Figure~\ref{four} shows a plot of the vevs of the odd and even four fermion operators as a function of the even site coupling $G$. The odd site
coupling is fixed at $g=0.01$. We use lattices of size $L=8$ and $L=16$ and anti-periodic boundary conditions for the fermions in the (Euclidean) temporal direction. Notice that the two vevs differ by four orders of magnitude. 
We can extract expressions for these condensates from the fermion
propagator which is given by inverting the fermion matrix $M$ which is given in odd/even block form by
\begin{equation}
M=\left(\begin{array}{cc}G\Sigma_+&\eta_\mu D_\mu\\
\eta_\mu D_\mu & g\Sigma_-\end{array}\right)
\end{equation}
where $\Sigma_\pm=\sigma_\pm^A\Gamma_A$. This yields
\begin{equation}
   <\chi\chi>= \left(\begin{array}{cc}
    A & B \\
    B & C\end{array}\right)
\end{equation}
where
\begin{align}
   \left\langle \chi_+\chi_+\right\rangle&=A=g\Phi_+\\\nonumber
   \left\langle \chi_-\chi_-\right\rangle&=C=G\Phi_-\\\nonumber
   \left\langle \chi_+\chi_-\right\rangle&=B=\Sigma_+^{-1}\eta_\mu D_\mu\Phi_-
\end{align}
where
\begin{equation}
    \Phi_\pm=\left( x\Sigma_\pm -\eta_\mu D_\mu \Sigma^{-1}_\mp \eta_\mu D_\mu \right)^{-1}
\end{equation}
For $x<<1$ this yields 
\begin{align}
\label{theory}
<(\chi_+\Gamma\chi_+)^2>&=g^2\\\nonumber
<(\chi_-\Gamma\chi_-)^2>&=G^2
\end{align}
The dashed line shows a fit to
the data assuming the coupling constant dependence shown in eqn.~\ref{theory}. In the same limit the
off-diagonal propagator takes the free field form 
\begin{equation}
<\chi_+\chi_->=\frac{\eta_\mu D_\mu}{\Delta\mu^2}    
\end{equation}

\begin{figure}[htb]\centering
    \includegraphics[width=0.4\textwidth]{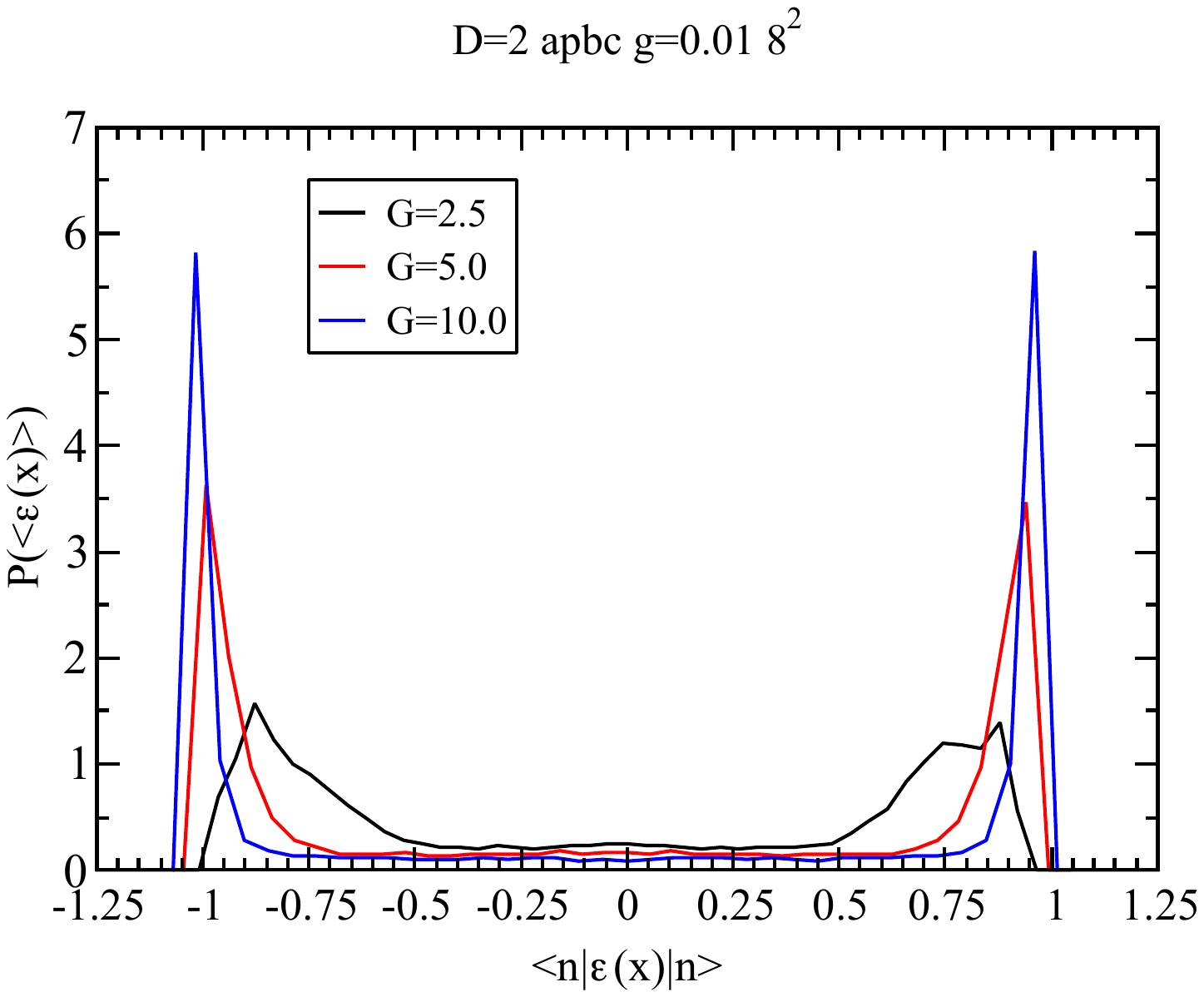}
    \caption{Histogram of $<\epsilon(x)>$ for $L=8$ and $G=5.0$, $g=0.01$}
    \label{hist}
\end{figure}

To gain more understanding of the implications of this asymmetric
four fermion condensate it is useful to look at the spectrum of the fermion operator.
Figure~\ref{eigen} shows a plot of the 
matrix element of the site parity operator $\sum_x \phi_n^2(x) \epsilon(x)$
for each eigenstate $\phi_n(x)$ of the fermion operator as a function of its
corresponding eigenvalue $\lambda_n$ for $L=8$ at $G=5.0$. The symmetry in the plot just reflects the fact that the fermion eigenvalues come in equal 
and opposite pairs as expected for an antisymmetric matrix.
Clearly the low lying
eigenstates have odd parity $\epsilon(x)\sim -1$ while the states with large
eigenvalue have even parity $\epsilon(x)\sim +1$. 
This conclusion is reinforced if we look at the histogram of the values of $\epsilon(x)$ shown in figure~\ref{hist}
which shows that the distribution of values of $<n|\epsilon(x)|n>$ cluster increasingly around $\pm 1$ as $G$ increases.
As $G\to 0$ it is easy to verify that
$<n | \epsilon(x)|n>=0$ for all modes i.e the double peaks merge back into a single peak at the origin.
This is to be expected  since
$\phi_n(x)$ and $\epsilon(x)\phi_n(x)$ are orthogonal eigenvectors of the fermion matrix at $G=0$ which follows from the
fact that $\epsilon$ anticommutes with the staggered fermion operator.

\begin{figure}[htb]\centering
    \includegraphics[width=0.4\textwidth]{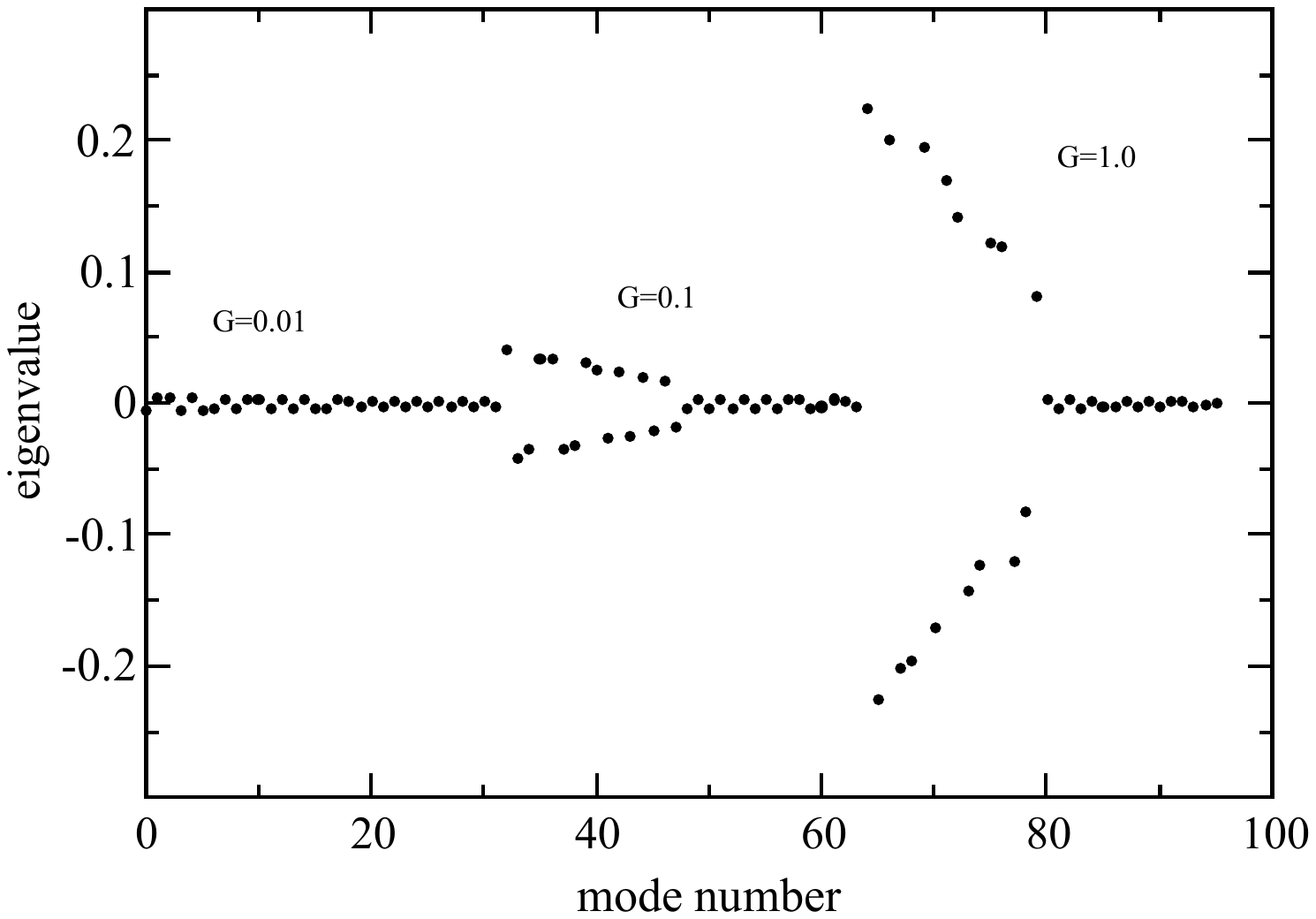}
    \caption{First 32 eigenvalues of the fermion operator for $G=0.01$, $G=0.1$ and $G=1.0$ for $8^2$ lattice
    at $g=0.01$ and pbc}
    \label{mode}
\end{figure}

It is instructive to also examine the lowest eigenvalues of the fermion operator $M$ as the coupling $G$ is varied.
This is shown in fig.~\ref{mode} which plots the lowest 32 eigenvalues of $M$ for three different values of the
coupling $G$ on a $8^2$ lattice with periodic boundary conditions and $g=0.01$.
Each eigenvalue is given a unique label in the range $0-95$ with the first 32 eigenvalues
corresponding to
$G=0.01$, the next 32 to $G=0.1$ and the final 32 to $G=1.0$. Notice again that the eigenvalues come in $\lambda,-\lambda$ pairs as expected from antisymmetry of $M$.
At the smallest coupling we see that all 32 eigenvalues lie close to zero. This is to be expected since eight
flavors of two dimensional massless real staggered fermions should have $4\times 8=32$ exact zeroes.
However as $G$ is increases half of these would be zero modes are progressively lifted
away from the origin because of the strong Yukawa interaction. Indeed if we are successful
at gapping half of the modes we would expect only sixteen would-be zero modes to remain  corresponding to 
eight Weyl fermions in two dimensions. This is exactly what we see happens in this plot as  $G$ increases. 

\begin{figure}[htb]\centering
    \includegraphics[width=0.4\textwidth]{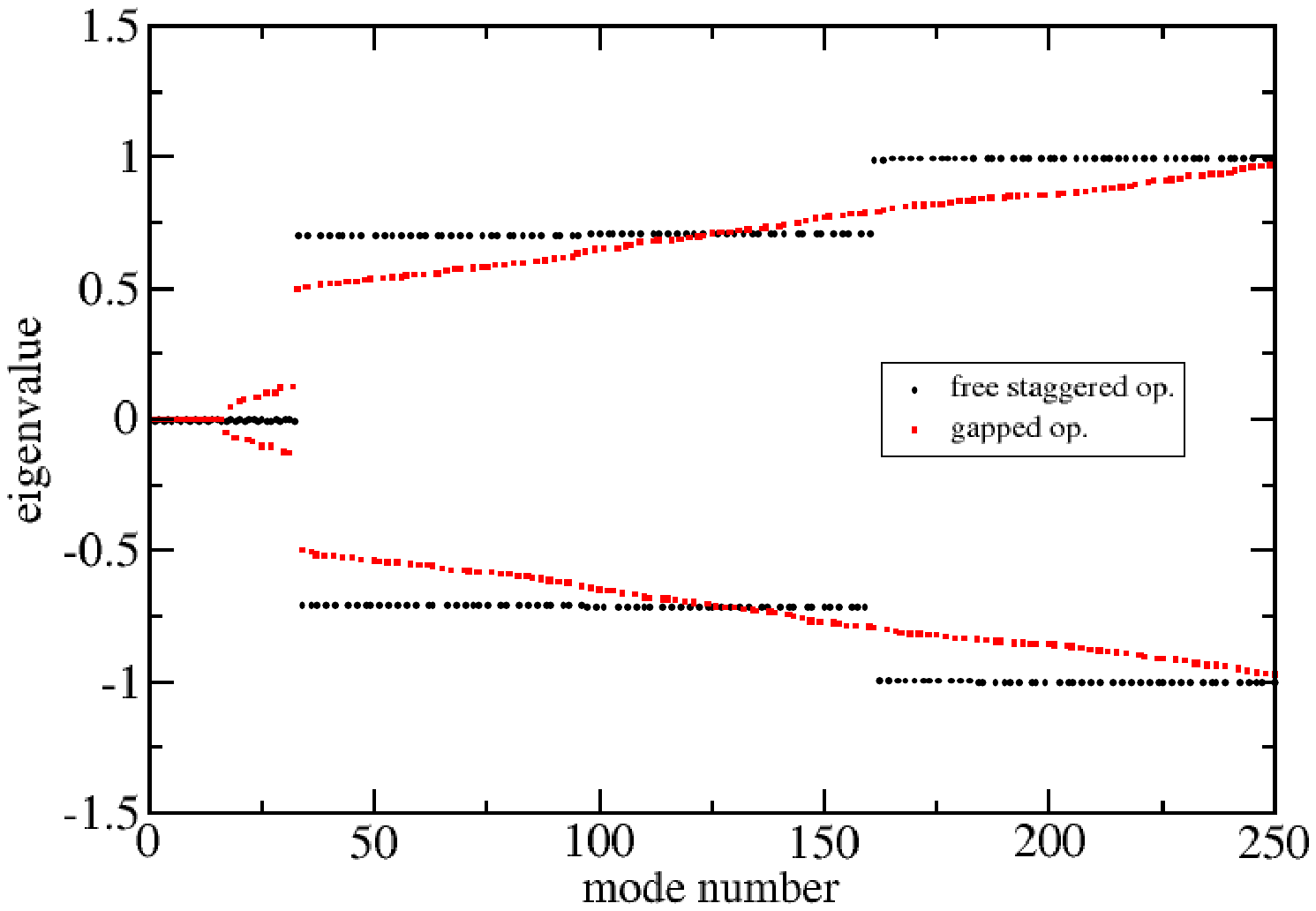}
    \caption{Spectrum of $8^2$ periodic lattice at $g=0.001$ and $G=0.5$}
    \label{spec}
\end{figure}

Of course we are most interested in whether the low lying non-zero eigenvalues
of the gapped theory correspond to what one would expect for a Weyl fermion. 
Figure~\ref{spec} shows the lowest 256 eigenvalues of $M$ at $G=0.5$ and $g=0.001$ for an $8^2$ lattice with periodic boundary conditions.
For comparison we also plot  the spectrum of the free
reduced staggered fermion operator for the same lattice.
While we see a rough matching of
the spectrum to that of a free fermion the correspondence is not perfect which presumably 
reflects the effects of the
fluctuating scalar field which has a small but non-zero coupling to the odd parity fermions.
However, in any case, one should be careful in reading too much from this comparison; the
continuum Weyl fields are to be constructed from the odd parity staggered
fields in each elementary square of the lattice and hence reside
on a lattice with twice the original lattice spacing. It is the spectrum of the effective quadratic
operator coupling these odd parity fields on this coarser lattice that is of primary interest.
Construction of the effective quadratic operator
governing these Weyl fermions on the coarser lattice
is a future goal of our work and will be necessary to be sure the theory flows to
a chiral theory in the continuum. 

The other key issue we would like to understand better is whether the system
develops a fermion bilinear condensate. If it does the
entire approach will fail.  To see this go back to the form of the action written in eqn.~\ref{rescaled} and replace $\sigma_A$ by a constant.
Operating with $\eta_\mu  D_\mu$ on the  equation of motion for the light field yields \begin{equation}
D_\mu^2 \chi_--\sigma^2\chi_-=0\end{equation} 
which shows that the $\chi_-$ field picks up a $\mu$ independent mass which is equal to that for the $\chi_+$ field and
indicates that the continuum limit will correspond to a Dirac fermion.

\begin{figure}[htb]\centering
    \includegraphics[width=0.4\textwidth]{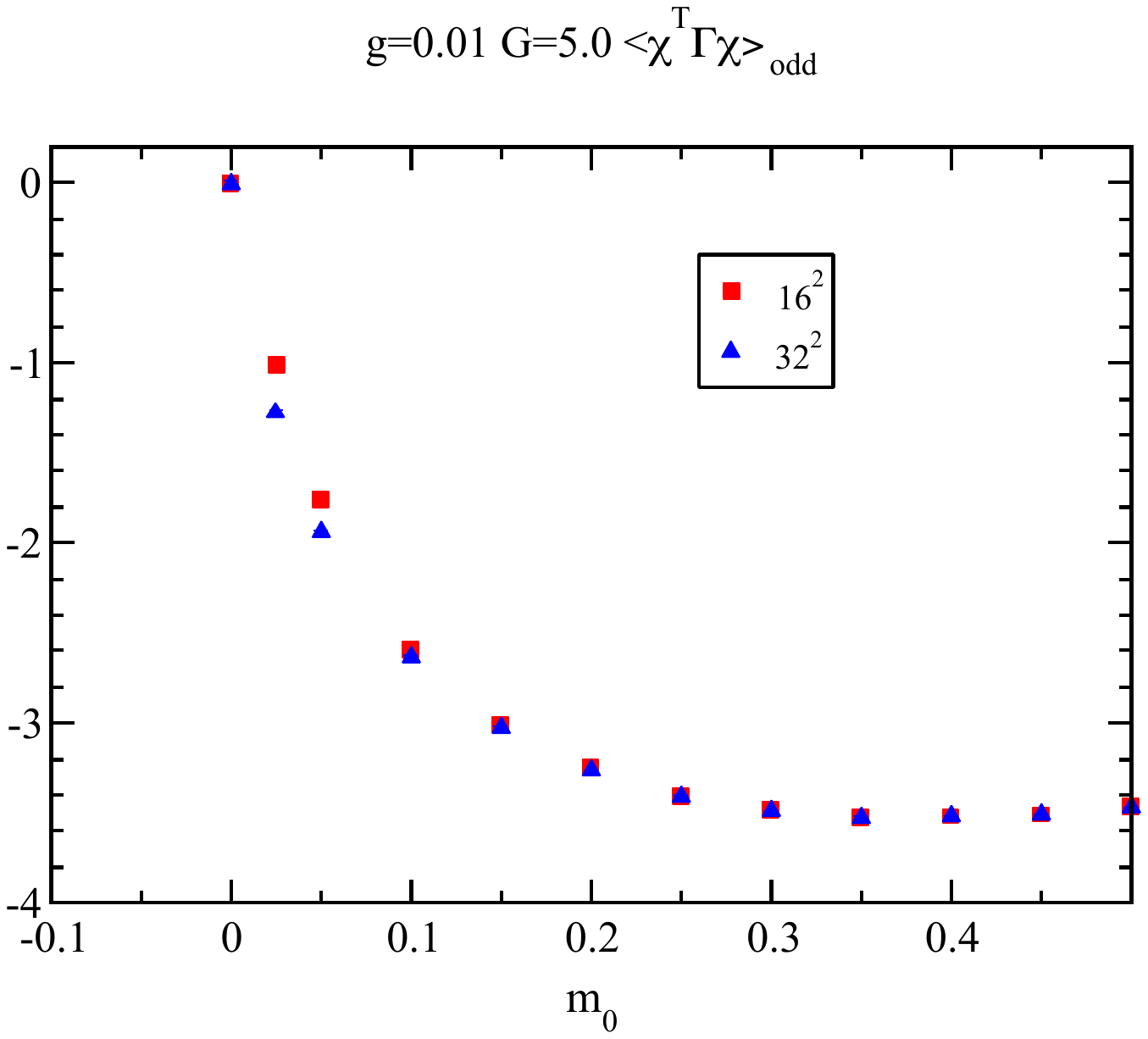}
    \caption{$<(\chi^T\Gamma_0\chi)>$ vs $m_0$ for $L=16$ and $L=32$}
    \label{bi_odd}
\end{figure}

\begin{figure}[htb]\centering
    \includegraphics[width=0.4\textwidth]{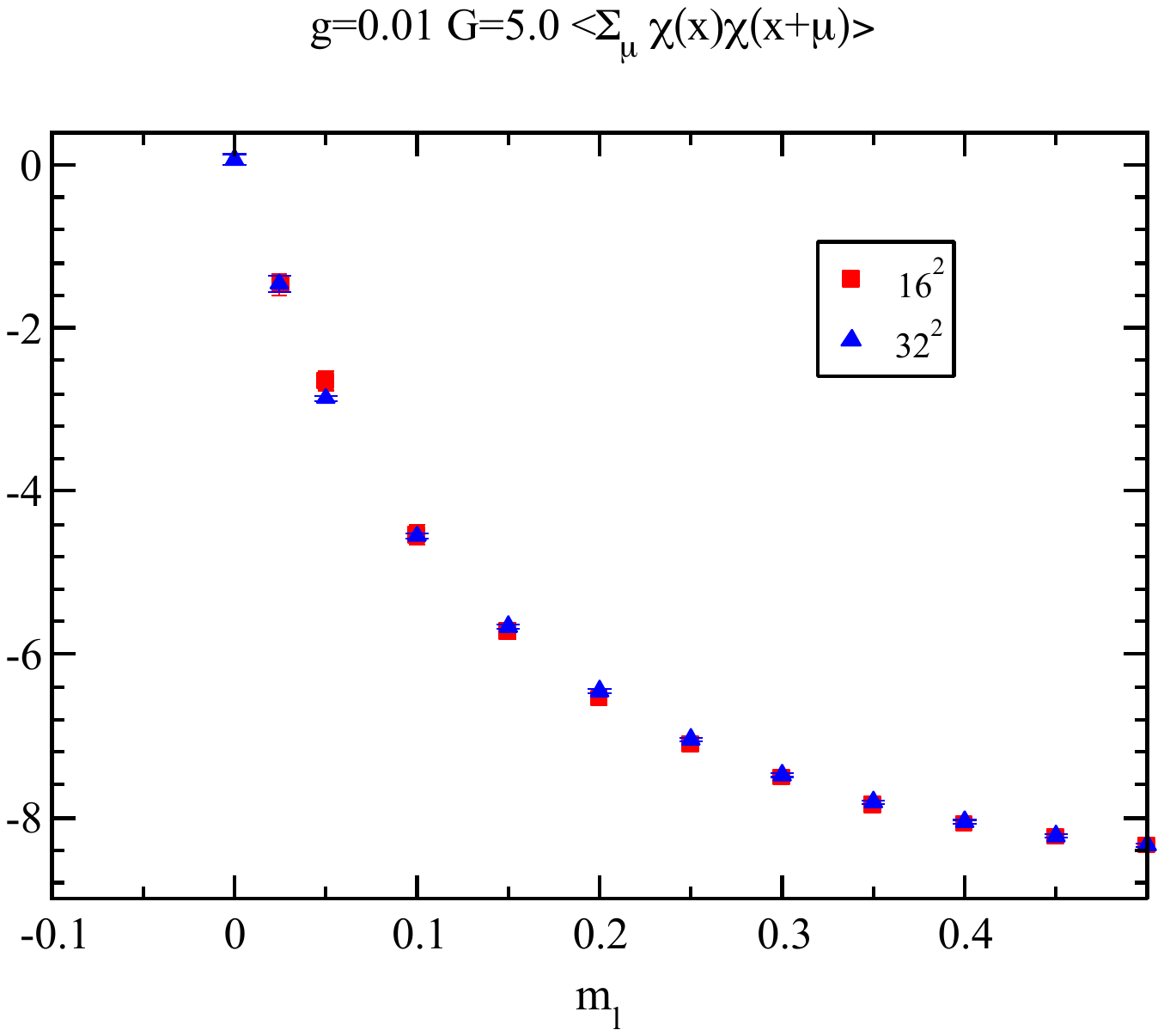}
    \caption{$\sum_\mu<\chi^T(x)\chi(x+\mu)\epsilon(x)\xi_\mu(x))>$ vs $m_1$ for $L=16$ and $L=32$}
    \label{lcond}
\end{figure}

To test for this we have focused our attention on
two particular bilinear operators - the site operator
\begin{equation}
    O_1=m_0<\chi^T\Gamma_0\chi>
\end{equation}
and the link operator  \cite{vandenDoel:1983mf,Golterman:1984cy}
\begin{equation}
    O_2=m_1\sum_\mu \chi^T(x)\chi(x+\mu)\epsilon(x)\xi_\mu(x)
\end{equation}
A vev for the former would spontaneously break ${\rm Spin}(7)$ and the $Z_4$ 
symmetry while a vev for the latter would explicitly couple the even and
odd sectors. 
To test for these scenarios we have added explicit sources to the action
and measured the vevs for several lattice volumes as the sources are sent to zero.
In systems exhibiting spontaneous symmetry the corresponding vev develops
a strong volume dependence as the source is sent to zero allowing for a non-zero 
vev to survive in the thermodynamic limit. However, as is clear in figure~\ref{bi_odd}
and figure~\ref{lcond} which employ $G=5.0$ and $g=0.01$ no such strong volume dependence is seen 
for either bilinear and the
vevs go smoothly to zero with vanishing external source. We thus conclude
that there is no evidence for the formation of non-vanishing fermion bilinears at least
in the two dimensional model.

It is perhaps useful to pause at the point to reiterate the basic strategy of this approach.
In domain wall fermion approaches to chiral fermions one is able to separate the chiral modes in an extra fifth dimension. Here, in contrast, one
uses Yukawa interactions to
separate the eigenstates of the parity operator $\epsilon(x)$  
according to their eigenvalue (essentially separating them in momentum space). Of course for the 
lattice theory to target a chiral theory in the continuum limit requires that there be a correlation between
$\epsilon(x)$ and $\gamma_5$. 
For regular staggered fermions containing both $\chi$ and $\overline{\chi}$ at each site 
there is no simple correspondence between the site parity operator $\epsilon(x)$ and the continuum chirality
operator $\gamma_5$ -  lattice fields with a given value of $\epsilon(x)$ give rise to both left and right handed
continuum fermions. However, the correspondence becomes closer if one restricts to real staggered
fields where the discussion in section~\ref{chiral} shows that $\epsilon(x)$ becomes a proxy for
$\gamma_5$ at non-zero lattice spacing. However, even in that case, the continuum limit of the non-interacting
theory is still vector-like
since it contains equal numbers of even and odd lattice fields leading to equal numbers of left and right
handed continuum fermions. If we now switch on $G$ it is no longer true that $\epsilon(x)$ is a proxy for $\gamma_5$
for  the even parity fields. But it continues to be true for the odd parity fields since they
are not subject to the Yukawa interaction. If symmetric mass generation occurs and the even parity modes are
removed from the low energy spectrum and given a mass of order the
cut-off we will be left with a theory that targets fermions of fixed chirality in the continuum limit. This is the
strategy being pursued in this paper. 

\section{Summary and Prospects}

In this paper we have described a new lattice fermion mirror model 
which employs reduced staggered fermions transforming in the
real eight dimensional chiral representation of a ${\rm Spin}(7)$ symmetry group.
Interactions are introduced via Yukawa interactions of Fidkowski-Kitaev type on even
parity lattice sites which we argue are capable
of generating mass for half of the lattice fermions without producing
symmetry breaking bilinear fermion condensates. Gapping fermions
without breaking symmetries is called symmetric mass generation.

If these Yukawa couplings are able to generate cut-off scale masses for the even parity modes 
we argue that the remaining odd parity modes can be reorganized, in four dimensions, 
into sixteen free Weyl fermions in the
continuum limit. This number of Weyl fermions is precisely what is needed to cancel off a discrete
spin-$Z_4$ anomaly in the continuum theory and indeed the vanishing of this anomaly is a necessary
condition to decouple the mirror fermions
via symmetric mass generation. Indeed, with hindsight, the failure to cancel off
these discrete anomalies was one of
the primary reasons that many early efforts to construct lattice mirror models failed.
In the appendix we give an argument that this fermion counting is also consistent with
the cancellation of a novel gravitational anomaly that is unique to a generalization of staggered fermions 
propagating on lattices of arbitrary topology.

We show numerical evidence in favor of this scenario from 
simulations in two dimensions. The structure of
this interaction, the arguments for symmetric mass generation, and the chiral properties of the
continuum limit are identical to the case of four dimensions. In two dimensions a reduced staggered
field contains four real degrees of freedom in a unit cell and hence yields a single Dirac fermion in the
continuum limit.   Since the model contains eight copies of this field the continuum limit  of the free theory
can be rewritten in terms of eight left and eight right handed Weyl fermions. If symmetric mass generation occurs
and Lorentz invariance is restored,
the gapped theory will possess just eight left Weyl fermions in the continuum limit -
a number which matches that required for cancellation of another discrete anomaly arising from chiral fermion parity
\cite{Ryu:2012he,Qi:2013dsa,Garcia-Etxebarria:2018ajm,Razamat:2020kyf}.
For large Yukawa coupling $G$ our simulations shows that the low lying modes of
the fermion operator do indeed have odd parity while the even parity modes are heavy. Furthermore, we
observe that 
the number of zero modes of the fermion operator agrees with what is expected
for 8 copies of a free Weyl fermion. Assuming Lorentz invariance is restored these ingredients suggest that
free Weyl fermions will indeed be recovered in the continuum limit.

It also seems likely that a similar construction may work in three dimensions.
In this case the unit cube on the lattice contains eight real staggered fields. Gapping say the even site parity fields would leave four light fields which can
be assembled into two Majorana fermions in the continuum limit. Since these transform in an eight dimensional
representation the continuum limit would
then contain sixteen Majorana fermions again consistent with 
the expectation from continuum time reversal invariance.

In this paper the 
the ${\rm Spin}(7)$ symmetry is a global symmetry. However, it can be straightforwardly
gauged to yield a chiral lattice gauge theory in the continuum limit. One
merely replaces the symmetric difference operator by the appropriate
gauge covariant difference operator acting on fermions in the eight
dimensional representation of ${\rm Spin}(7)$. 
\begin{equation}
D_{\rm cov}\chi(x)=\frac{1}{2}\left(U_\mu(x)\chi(x+\mu)-U^\dagger(x-\mu)\chi(x-\mu)\right)
\end{equation}
where $U_\mu(x)=e^{i\omega_{AB}\left[\Gamma_A,\Gamma_B\right]}$ are the gauge links for the real eight
dimensional chiral representation of ${\rm Spin}(7)$.
The interaction terms
being purely local are already ${\rm Spin}(7)$ gauge invariant. 

Of course, the question arises as to how one should take the continuum limit. 
Naively one would expect to have to search for a continuous phase transition in the theory as was done for
the vector-like case. However, while such a phase transition
would presumably be needed to take a continuum limit for the strongly interacting sector, 
it does not appear necessary if the only
fields of interest lie in the non-interacting odd parity sector.
Indeed it appears that keeping $x=gG<<1$ fixed 
while $y=\frac{g}{G}\to 0$ may be sufficient. 
Since $G$ is naively an irrelevant coupling in four dimensions with mass dimension minus one
one could imagine scaling its bare coupling $G\sim L$ as $L\to\infty$ to retain a cut-off
scale four
fermion condensate for the even parity fields as the lattice size is increased.
To keep $x$ fixed one would then 
simultaneously scale $g\sim \frac{1}{L}$. We are currently exploring this limit in more detail. 

Some caveats are in order. 
We have assumed Lorentz invariance is restored in the continuum limit in order to identify the chirality and flavor representation of the  fermions arising from the underlying staggered lattice fields. More specifically
we have assumed that the low lying lattice modes organize themselves into Weyl spinors as the 
lattice spacing approaches zero.  While this is consistent with our counting
of degrees of freedom and the constraints from discrete anomalies
it is a stronger requirement than simply the restoration of rotational
invariance. It should be checked in future numerical work.
In addition, while we have observed no sign problems for the simulations
reported in this paper a sign problem will likely return once
one gauges the ${\rm Spin}(7)$ flavor symmetry and/or takes the continuum limit.
Finally, while we have see no evidence for the formation of
off-site fermion bilinear condensates in two dimensions, this must
be carefully checked in four dimensions.

\acknowledgments
This work was supported by the US Department of Energy (DOE), Office of Science, Office of High Energy Physics under Award Number {DE-SC0009998}. 
The author would like to thank Nouman Butt and Goksu can Toga for discussions and David Tong for patiently educating me about discrete anomalies.
\bibliography{chiral}

\appendix
\section{\label{sec:anom}Anomalies and Staggered Fermions}

There is another way of understanding why the lattice theories studied in this
paper are subject to strong constraints on their fermion content. To exhibit these constraints it is
necessary to generalize staggered fermions to lattices with non trivial topology. The starting point 
is to recognize that staggered fermions can be thought of as a particular
discretization of \KD fermions -- albeit one suitable only for
regular toroidal lattices \cite{Banks:1982iq}. 

The \KD equation offers an alternative to the Dirac equation
for curved spaces and reduces to the latter for flat spaces where it describes multiples of Dirac fermions (four flavors
of Dirac fermion in four dimensions). It takes the form
\begin{equation}
\left(d-d^\dagger+m\right)\Omega=0\end{equation}
where $\Omega$ is a collection of antisymmetric tensor fields over the space and $d$, $d^\dagger$ are the exterior
derivative and its adjoint.  In flat space the tensor fields are given in terms of the spinor components of this set of
Dirac fermions and the antisymmetry of the forms reflects the antisymmetric properties of the Clifford algebra
associated to the Dirac gamma matrices.

A general procedure for discretizing the \KD equation on simplicial lattices 
was given in \cite{Rabin:1981qj}. First, the antisymmetric component fields in $\Omega$ are placed on p-simplices in an (oriented) triangulation of the space. Thus lattice fields are placed on sites (0-simplices), links (1-simplices), triangles (2-simplices) etc. Next, a discrete boundary
operator $\partial$ is defined whose action on some $p$-simplex decomposes it into an oriented list of its $(p-1)$-simplex boundary components. It can hence be used to create
a map between $p$-simplex fields and $(p-1)$-simplex fields. 
In fact this boundary operator is the discrete analog of the adjoint of the
exterior derivative. 
A similar lattice operator - the co-boundary operator $\overline{\partial}$ when acting on a $p$-simplex
returns an oriented list of $(p+1)$-simplices that contain that $p$-simplex in their boundary. The co-boundary operator
replaces the exterior derivative and
yields a map between  $p$-simplex fields and $(p+1)$-simplex fields.
The massless discrete \KD equation then takes the form
\begin{equation}
    (\partial-\overline{\partial})\Omega=0
\end{equation}
Solutions of this equation go smoothly over into their continuum cousins as the lattice spacing is reduced - there is
no additional fermion doubling.

In \cite{Catterall:2018lkj} it was shown that the corresponding \KD 
action- both continuum and discrete - is invariant under a $U(1)$ symmetry which distinguishes
tensor fields with an even or odd number of indices. It acts
as 
\begin{align}
\Omega&\to e^{i\alpha\Gamma}\Omega\\
\overline{\Omega}&\to \overline{\Omega}e^{i\alpha\Gamma}
\end{align}
where $\Gamma$ anticommutes with the lattice \KD operator and takes the values $\pm 1$ on even and odd
forms.
This transformation is the analog of  $e^{i\epsilon(x)\alpha}$ for
staggered fermions on a torus but works on an arbitrary random
triangulation of any topology. However, as was shown in \cite{Catterall:2018lkj} 
this symmetry is anomalous with the resulting partition function
transforming by an overall phase  $e^{i\alpha\chi}$ depending only on the Euler character of the triangulation which is
given by
\beq\chi=N_0-N_1+\ldots \left(-1\right)^pN_p+\ldots +\left(-1\right)^DN_D
\eeq
Notice that this result holds equally well in the continuum for \KD fermions and shows that the anomaly
can be computed exactly in the lattice theory since it depends only on the topology of the background space which
can be captured exactly by the lattice. It is an example of an anomaly which {\it does not} require the presence
of an infinite
number of degrees of freedom.

Applying this result
to the sphere $S^{2n}$, which we can think of as representing
a compactification of $R^{2n}$, one finds that this $U(1)$ symmetry is broken to
$Z_{4}$ in even dimensions. Thus, while a fermion bilinear operator is prohibited by this symmetry,
it is possible to introduce four fermion operators. If want to retain Lorentz invariance
we are forced to consider theories with at least four flavors of \KD field. If we restrict to real
representations these four complex \KD field can be decomposed into eight real \KD fields which are
equivalent to eight reduced staggered fermions on a torus. This is precisely the field content we argued was necessary to formulate a mirror model capable of symmetric mass generation and yields a theory after gapping with
exactly eight and sixteen massless Majorana fermions in two and four dimensions respectively. This suggests that there
is a connection between the spin-$Z_4$ symmetry of the continuum and this $Z_4$ symmetry of \KD fermions.

\section{\label{rhmc} Numerical methods}
If we integrate out the fermions we obtain a Pfaffian ${\rm Pf}(M)$. Provided this is positive definite we
can replace it by ${\rm det}\left(M^\dagger M\right)^{\frac{1}{4}}$. In practice the latter is generated
by integrating over a set of commuting pseudofermion fields with action
\begin{equation}
S_{\rm PF}= \phi^T\left(M^\dagger M\right)^{-\frac{1}{4}}\phi=
\sum_i^N \phi^T \left[\frac{\alpha_i}{M^\dagger M+\beta_i}\right]\phi
\end{equation}
where $\phi^a(x)$ is vector indexed by the lattice sites and ${\rm Spin}(7)$ label.
The second expression represents a rational fraction approximation to the fractional power of the matrix.
The coefficients $\alpha_i$ and $\beta_i$ are determined for a given number of terms $N$ by the remez
algorithm - see \cite{Clark:2004cp}. In practice we set $N=18$ and tolerate a relative error of $10^{-8}$ over the eigenvalue
interval $0.00000001-1000.0$.
We use a multi-timestep Omelyan integrator to generate an auxiliary classical dynamics  that is used to
sample the partition function of the system where the configurations are subject to a Metropolis test after each
molecular dynamics trajectory. The time consuming part of this evolution corresponds to the calculation of
the force terms arising from the pseudofermion action, For these we use a multi-shift CG solver. 
We have implemented a parallelized version of this algorithm by using the MILC communication libraries
and the code runs efficiently on clusters with dedicated networking.

Our results derive from ensembles of 2000 configurations for a given lattice size and
set of couplings with measurements taken every 10 sweeps. Errors are assessed as usual by a jackknife procedure
and a set of $Z_2$ stochastic sources are used to estimate condensates.

\end{document}